# Gedanken Tests for Correlated Michelson Interferometers

One- interferometer tests and two-interferometer tests: the role of cosmologically-implemented models including Poincaré particles

Orchidea Maria Lecian[1, 2*]

[1]Comenius University in Bratislava, Faculty of Mathematics, Physics and Informatics, Department of Theoretical Physics and Physics Education- KTFDF, Mlynskà Dolina F2, 842 48, Bratislava, Slovakia
[2]Sapienza University of Rome, Faculty of Civil, Constructional and Environmental Engineering, DICEA, Via Eudossiana, 18- 00184 Rome, Italy
[*]Corresponding author

*Abstract*—The features of correlated Michelson interferometers are for describing the analysis of Einsteinian spacetime models, and the quantum geometries pertinent with descriptions of GR compatible with particle Physics. Such apparati allow for the spectral decomposition of fractional Planck-scale displacements correlations and fractional Planck-time-interval correlations for kinematical investigations in particle Physics on emerging Minkowski background, and for models which admit GR as a limit after cosmological implementations for Poincaré particles content.

## I. Introduction

Two Michelson interferometers system test allow to constrain anomalous effects of decoherence between two spacelike-separated geodesics, for which the spectral analysis of the quantum-noise correlation of four-dimensional position of bodies (i.e. the mirrors at which the angular uncertainty of geodesics is calculated) is calculated with (suitably-normalized) power smaller than a Plank time $t_{Pl}$, for which its strain noise power spectral density for the fractional position displacements can e able to removes any correlation within high-frequency cut-offs an low-frequency cut-offs [1].

The present spectral analysis simulation is aimed at comparing the same terms possibly contributed by ultra-energetic strongly Lorentz violating cosmic ray events[2], and compatible with modifications of GR[3] (for which adaptations of the Wigner-Bargman O(3) can be factly considered) at the linearized (low-energy) for the phase space (i.e. also within modified dispersion relations) [4]. The most extreme case, for which Poincaré symmetry is maintained only in the direction of only one interferometer arm [5], is also considered: for this, different manifestations of the Fifth Force (FF)[6] and the limits of non-Riemannian structures [7] for the available spectral definition are comprehended within Poincar'e symmetry and a coherent cosmological evolution.

Alternative implementations [8] compatible with the same spectral analysis are to be considered within the evaluated variances.

## II. Theoretical remarks and experimental decsription

### A. Planck Length Resolutions for the Spectral Analysis

The beam splitters characteristics of the apparati [1] reduce the susceptibility to of lensing effects, and the end-mirrors transmission optics allows for 'ad hoc' alignment procedures. The calibration for all higher frequencies is safely assured by the knowledge of the electronics only , while the simultaneous sampling of lower frequencies can be resolved by system control techniques [9] for the additional electronic response terms. The placements of the beam splitters would allow to appreciate space (time) displacements fractional wrt the Planckian length $L_{Pl}$, which allow for a spectral analysis of the strain noise correlation at samples timed for time intervals smaller than the Planck time $t_{Pl}$.

### B. Spectral Analysis in General-Relativity Settings

In the present work we are mostly interested in those cases for which generic modified dispersion relations $E^2 = F(\mathbf{p})$, with $\mathbf{p}$ the three-momentum of a particle, can be specified by recasting them under more restrictive hypotheses, under which the modification to the dispersion relations can be considered as a modification term to the (low-energy limit[2]) Minkowskian expression $E^2 = m^2 + |\mathbf{p}|^2$, i.e. such that

$$E^2 = m^2 + |\mathbf{p}|^2 + f_\rho(E_\rho, m_\rho, \mathbf{p}_\rho; C_r(\rho); L_q)$$

where the modification to the dispersion relation (1) can be expressed as a function $f$, which can be chosen to depend on the energy of the particle, its momentum, its position, other properties, such as quantum symmetries which can be further described by a set of $n_C$ parameters $C_r(\rho)$, $r = 1, ..., n_C$; such a modification of the dispersion relation an furthermore be indicated to depend also on particular m energy scales $L_\varsigma$ (or length scales), $\varsigma = 1, ..., m_\varsigma$, at which the modification(s) would appear phenomenologically more relevantly suited to describe modifications to the dispersion relations which depend only onthe matter content of the spacetime, or on the features of the space(-time) distances, or specifically on both definition within Relativistic Poincaré symmetry.

Among these settings, there appear modelizations implying a maximum momentum but not a maximum energy, or an upper bound for the energy but not on the momenta as a consequence of the choice of specific lenghts or cut-off's [10].

## III. OPTICAL POWER VARIATIONS

### A. Modified Dispersion Relations Which Admit GR in the Linearized Limit

Classes of models are analyzed, whose kynematics analysis allows for the same (tangent) phase space after the linearized regime both for the modified Minkowski model and for the modified General-Relativity phase- space model.

It is possible to consider [4]

$$\omega^2 \gamma = k^2 + \xi k^n, \quad (2a)$$

$$E_e^2(p) = m_e^2 + p^2 + \eta\, p^n, \quad (2b)$$

$$E_i^2(p) = m_i^2 + p^2 + \eta\, \varepsilon_i\, p^n, \quad (2c)$$

for which the dispersion relations specify differently for photons (2.1a), other particles (such as electrons) (2.1b), and neutrinos $\nu$, specified with their generation (composition) parameter $\varepsilon_i$.

The kinematics of the phenomena $\gamma \rightarrow \nu_i \nu_j$, and the kinematically analogous $j$ $a nit\nu$ proces(es), admit the same first order expansion for the threshold energy required, $K\nu$, by setting $\varepsilon_i = 1$ for any $\nu$ generation and helicity state. Such a schemtiazion also applies ot the tangent space, which is suited for comparison with models which admits General Relativity as a possible (low energy) limit.

The considered model has the peculiarity to admit the following threshold theorem: if the threshold momentum for a photon/particle process is a strictly increasing monotonic function, the momenta of the 'out' ($\infty$) states are parallel.

The kynematical schematization will be useful in the following, for the analysis of the model for not including the composition-dependence description in case of the hypothesis of the Fifth force.

### B. Contributions Comprarable with GR

A difference in the integration of the optical power Shear-noise correlation functions for the two interferometers can therefore reveal the occurrence of such phenomena for the parallel out states and the antiparallel ones, which do not imply any angular deviations (experimental indeterminacies) for the corresponding geodesics by comparing the difference in the optical powers of the two beam-splitters after photon-rate integration, in the case of distinguishing [12] each (of the two) results of the expectation value of the (one sided) power spectrum for a decrease $-P(\xi)$ of the optical power spectrum both in the frequency $\omega$ and in the average photon rate, $\dot{N}$ wrt the proper time of the Minkowski interferometer emerging space., for whose evolution one obtains

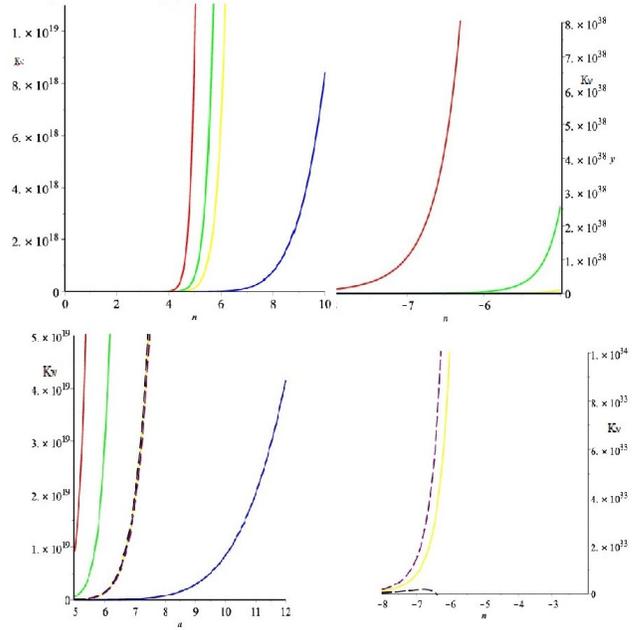

FIGURE I. NUMERICAL SIMULATIONS: THE FRACTIONAL SPECRAL ANALYSES FORTHE AVAILABLE PHASE SPACE FOR THE CONISDERED PROCESSES

The available phase space for the particle/antiparticle processes involving $\nu$s is schematized in Figure 1 for positive n and in Figure 2 for negative n. The thresholds for the momenta $K\nu$ as a function of the momenta, for different orders of the expansion, with n positive. The chosen momenta are $k\nu = 0.01$ (green), $K\nu = 1$ (red), $K\nu = 10$ (yellow), $K\nu = 100000$ (blue).

In Figure 3, the $\nu$ mass indeterminacy (6) allows to gain insight on the free parameter n for the maximum standard deviation for the normalized threshold function $K\nu$, been plotted (black dashed line, purple dashed line) for $K\nu = 100$ (solid yellow line) for the muonic $\nu$'s for a threshold momentum for summed masses summed masses $m_1 + m_2 \simeq 30$ for positive n.

For negative n (solid, yellow line) in Figure 4 (for $K\nu = 100$). From Figure 4, it is straightforward to infer that the available phase space for models admitting a modified-phase space GR limit, such as [8], [10] and [15] is non-trivial with respect to the standard error propagation limits calculated for the results without angular deviation in Figure 3.

Figures (1), (2) and (3) can therefore be studied within the analysis (1a) by [12] while information about Figure (4) can be outlined from (1b) by [13].

$$2\pi \Delta P(\xi) = (\Delta(\dot{N}))\, \xi k + (1/2)\, \dot{N}\, \xi k^{n-1} \quad (3)$$

after requiring a flat power spectrum for the variation of $\dot{N}$ wrt the $\nu$ mass $|\delta\, \Delta(\dot{N})\, /\delta\, m_\nu\,|^2 \equiv 1/(t_{Pl}\, M_{Pl}^2)$ after the kinematical analysis of (2), for which

$$\Delta m_\nu = (c^{4/n}/(1+2^{2/n}))\Delta k_\nu (\xi/\hbar^2)^n, \qquad (4)$$

requiring $\Delta(\dot{N})=(\delta \Delta(\dot{N})/\delta m_\nu)\Delta m_\nu$.

*C. Fifth-Force Interactions*

Any Fifth-Force interaction within the product neutri-nos [6] independent of composition description and of gravitational interaction will be therefore depending on other quality/charge numbers of the particles, andon their distance. It will have the effect to modify the geodesics describing the Minkowsky spacetime by deviating the straight-line 4-dimensional propagation of the photon beams, according the Fifth-Force specifications. The FF potential can be parameterized as

$$V_5 = [\alpha_{\infty; i, j}\exp[M_{Pl} r/(\lambda R)]]/r, \qquad (5)$$

where $\alpha_{\infty; i, j}$ factorizes the Newton gravitational constant at $\infty$, $G_\infty$, and is a function of the Fifth-Force numbers $\alpha_i$ and $\alpha_k$ of the $\nu$, $\nu_i$ and the other particle considered, k, with R the distance on which the interaction is considered; this way, $\beta$ corresponds to the other coupling constant for the Fifth Force (normalized by the Planck Mass) connected to $\lambda$ by numerical values describing the interaction within this ranges in the constant $\alpha_{\infty;i,j}$: $\alpha$ can be therefore factorized as $\alpha \approx G_N\alpha_0\alpha_i\alpha_k$, with $\alpha_j$ the Fifth-Force numbers relative to the particles, and $\alpha_0$ a numerical constant useful for relating the range of the Fifth Force with the gravitational constant $G_N$.

The features of the coupling constant for the Fifth Force can be modelled by the request to consider only non-composition interactions: for the kinematical analysis of $\nu$'s in (2), FF interaction with the cosmic $\nu$ background, whose mass has been re-quested to have an upper bound of $m_\nu \leq 0.2$ eV by cosmological investigation of the Cosmic microwave back-ground [14] within the COBE experiment, excludes $\nu$-$\nu$-FF interactions [15].

Poincaré symmetry in Fifth-Force interaction also within non-Riemannian cosmology can be recast as tachyonic fields Y are not admitted to emit vacuum Cerenkov radiation [10] because of the non-renormalizability of the operator term $\Box Y$ even on Minkowski spacetime: their propagation is consistent with the Lorentz invariance only in one space direction: the little group O (2, 1), which admits only a one-dimensional representation

[5]: for cosmological purpoes, it [7] can therefore be compared with the operator Y without violating local Lorentz invariance; to assure Poincaré symmetry in curved spacetime, the little group be realized in only one space direction (say, the direction individuated by the interferometer arm) to keep the *o operatorY tamed in cosmological implementation.*

*D. Non-composition FF Effects*

For the particle content of this emerged space-time, the term $(1/L_{Pl})$ appears, also by considering composition effects for the Fifth force (terms arising from $\alpha_i$ and $\alpha_k$). Variation of the gravitational constant $G_\infty$ through variation of $\alpha_0$ would be brought the same spectral information by assuming Poincaré symmetry. The corresponding $\nu$-FF interaction coupling constants interaction coupling constants as $\alpha_{\infty;i,k}$, $\alpha_0$, $\alpha_1$, $\alpha_k$, and the FF interaction coupling constant $\beta$, as it will be factorized out also within a non Einsteinian cosmological implementation of the Poincaré symmetry only in one of the interferometer arm.

## IV. ANGULAR INDETERMINACIES

Two different phenomenological interpretation of the analysis of the angular uncertainties [13], after the decomposition of the angular resolution wrt only one or both the two different Michelson interferometers, when the shear correlation functions between the two interferometers is normalized according to the length of the interferometer, but its derivative(s) consisting in the spectral analysis admit terms different from $c^2 t_{Pl}$.

*Poincaré Symmetry from (linearized) Deformed Phase Space*

In the case the momentum thresholds for models as (2) are not a monotonically increasing function, the 'out' ($\infty$) states are not parallel, but will have undergone an angular deviation $\Delta(\theta_\xi)$, the difference in optical power spectrum evaluates by combinig $K_\nu$ with (5)

$$\Delta(\theta_\xi)=[2\pi\Delta P_\nu/\hbar L]^{1/2} \qquad (6)$$

*Poincaré Symmetry after Poincaré-violating Phenomena: Cosmological Implementation in Non-composition iFF Interaction*

After avoiding $\nu$-$\nu$-FF interaction, the interaction of $\nu$'s within the FF can therefore be studied, the interaction of $\nu$'s with other the fields (5) admitted in the interferemoter emerging space and in the other implementations admits a maximal angular displacement obtained by majorizing the integral of the potential $V_5(r)$ at the end of the interferometer arm length (r=L) as

$$\Delta(x_5)=(\alpha_{\infty; i, k})/(\alpha_0 \alpha_i \alpha_k)(1/L^5)\exp(-M_{PL}/\beta) \qquad (7)$$

The angular deviation for the photons $N(\theta_5) \leq \Delta(x_5)/L$ constrains the coupling constant as smaller than that of the gravitational interaction by requesting $10^{-5} \mu \leq \mu_\beta \leq \mu\, 10^{-2}$ in the decomposition for the angular resolution

$$\Delta(\theta_5) = [L/\Delta(x_5)]^{1/2}$$

, containing two different eigenvalues for the deviation Poincaré eigenstates, which contain the inverse Planck length $L_{Pl}$ at linear power in the denominator

$$\delta\Delta(x_5) = \Delta(x_5)\,\Delta\beta\,M_{PL}/\beta^2 \qquad (8)$$

ACKNOWLEDGMENTS

The work of OML is funded by the National Schlarship Programme of the Slovak Republic (NS'P) SAIA (Slovak Academic Information Agency) Grant for International University Research Academic Year 2017/2018.

OML would like to thank Prof. V. Balek for the discussions about other regions of the phase space of Ref. [4]. Warmest hospitality at Faculty of Mathematics, Physics and Informatics, Department of Theoretical Physics and Physics Education, Comenius University in Bratislava is heartfully thanked. The software Maple was used for the graphics.